\documentclass[aps,amsmath,amssymb,twocolumn,preprintnumbers,nofootinbib]{revtex4-1}
\usepackage{amsmath,amssymb}
\usepackage{graphicx}
\usepackage{multirow}
   \usepackage{physics} 
\usepackage{acronym}
\usepackage{hyperref}

\newcommand{\beq}{\begin{eqnarray}}% can be used as {equation} or  {eqnarray}
\newcommand{\eeq}{\end{eqnarray}}

\begin{document}

% * title page
\title{Second wave COVID-19 pandemics in Europe: A Temporal Playbook}

\author{Giacomo Cacciapaglia} 
\email{cacciapaglia@ipnl.in2p3.fr}
\affiliation{\mbox{Institut de Physique des deux Infinis de Lyon (IP2I),  UMR5822, CNRS/IN2P3, F-69622, Villeurbanne, France}}
\affiliation{\mbox{University of Lyon, Universit{\' e} Claude Bernard Lyon 1,  F-69001, Lyon, France}}

\author{Corentin Cot} 
\email{cot@ipnl.in2p3.fr}
\affiliation{\mbox{Institut de Physique des deux Infinis de Lyon (IP2I),  UMR5822, CNRS/IN2P3, F-69622, Villeurbanne, France}}
\affiliation{\mbox{University of Lyon, Universit{\' e} Claude Bernard Lyon 1,  F-69001, Lyon, France}}
%\affiliation{\mbox{Univ. Lyon, Universit{\' e} Claude Bernard Lyon 1, CNRS/IN2P3, UMR5822 IP2I, F-69622, Villeurbanne, France}}

   \author{Francesco Sannino}
\email{sannino@cp3.sdu.dk}
\affiliation{CP3-Origins \& the Danish Institute for Advanced Study. University of Southern Denmark. Campusvej 55, DK-5230 Odense, Denmark; \\~\\
\mbox{ Dipartimento di Fisica E. Pancini, Universit\`a di Napoli Federico II | INFN sezione di Napoli}\\ \mbox{Complesso Universitario di Monte S. Angelo Edificio 6, via Cintia, 80126 Napoli, Italy.}}

\begin{abstract}
A second wave pandemic constitutes an imminent threat to society, with a potentially immense toll in terms of human lives and a devastating economic impact. We employ the epidemic renormalisation group approach to pandemics, together with the first wave data for COVID-19, to efficiently simulate the dynamics of disease transmission and spreading across different European countries. The framework allows us to  model, not only inter and extra European border control effects, but also
%to take into account 
the impact of social distancing for each country. We perform statistical analyses averaging on different level of human interaction across Europe and with the rest of the world. Our results are neatly summarised as an animation reporting the time evolution of the first and second waves of the European COVID-19 pandemic.  Our temporal playbook of the second wave pandemic can be used by governments, financial markets, the industries and individual citizens, to efficiently time, prepare  and implement local and global measures.
  \end{abstract}
%\preprint{CP3-Origins-2019-36 DNRF90}

\maketitle

%\section{Introduction and framework}
 Pandemics are increasingly becoming a constant menace to the human race, with COVID-19 being the latest example. A second wave is creeping back in Europe and is poised to rage across the continent by fall 2020.  
 
 In this letter we provide a statistical analysis of the temporal evolution of the second wave of infected cases, with the impact for various European countries. To model the spreading, we employ the \emph{epidemic Renormalisation Group} (eRG) framework, recently developed in \cite{DellaMorte:2020wlc,Cacciapaglia:2020mjf}. It can be mapped \cite{Cacciapaglia:2020mjf,morte2020renormalisation} into a time-dependent compartmental model of the SIR type  \cite{Kermack:1927}. The Renormalisation Group approach \cite{Wilson:1971bg,Wilson:1971dh} has a long history in physics with impact from particle  to condensed matter physics and beyond.  Its application to epidemic dynamics is complementary to other approaches  \cite{LI2019566,ZHAN2018437,Perc_2017, WANG20151,WANG20161,Danby85,Brauer2019,Miller2012,Murray,Fisman2014,Pell2018}.   
 
 The eRG approach consists in a set of first order differential equations apt to describe the time-evolution of the infected cases in a specific isolated region. It has been extended in~\cite{Cacciapaglia:2020mjf} to include interactions among multiple regions of the world, without the need for powerful numerical simulations.
The set of equations  \cite{Cacciapaglia:2020mjf} reads
\beq
\frac{d \alpha_i}{d t} = \gamma_i \alpha_i \left( 1-\frac{\alpha_i}{a_i} \right) +  \sum_{j\neq i} \frac{k_{ij}}{n_{mi}} (e^{\alpha_j - \alpha_i}  -1)\,,
\label{pandemic-eq}
\eeq
where 
\begin{equation} 
\alpha_i(t) = \rm ln\ \mathcal{I}_i(t) \ ,
\end{equation} with $\mathcal{I}_i (t)$ being the total number of infected cases {\it per million} inhabitants for region $i$ and $\ln$ indicating its natural logarithm. 
These equations embody, within a small number of parameters, the pandemic spreading dynamics across coupled regions of the world via the temporal evolution of $\alpha_i$, which resembles the energy dependence of the interaction coupling appearing in fundamental interactions of particle physics.  

The first term of the right-hand side in \eqref{pandemic-eq} characterises the epidemic evolution within a given region of the world.  The infection rate   $\gamma_i$, measured in inverse weeks, is responsible for how quick the epidemic evolves in the $i$-th region.
Besides depending on the intrinsic virulent character of the epidemic, the size of $\gamma_i$ can be controlled via social-distancing measures, with a flatter epidemic curve associated to smaller $\gamma_i$. 
 It is well understood \cite{Kermack:1927} that epidemic diffusion curves generally lead to plateaus in the total number of infected cases at late times. This is encoded in the parameter $a_i$, equal to the natural logarithm (ln) of the total number of infected cases (per million) at the end of the epidemic wave. 

\begin{figure}[tbh!]
\begin{center}
\includegraphics[width=8cm]{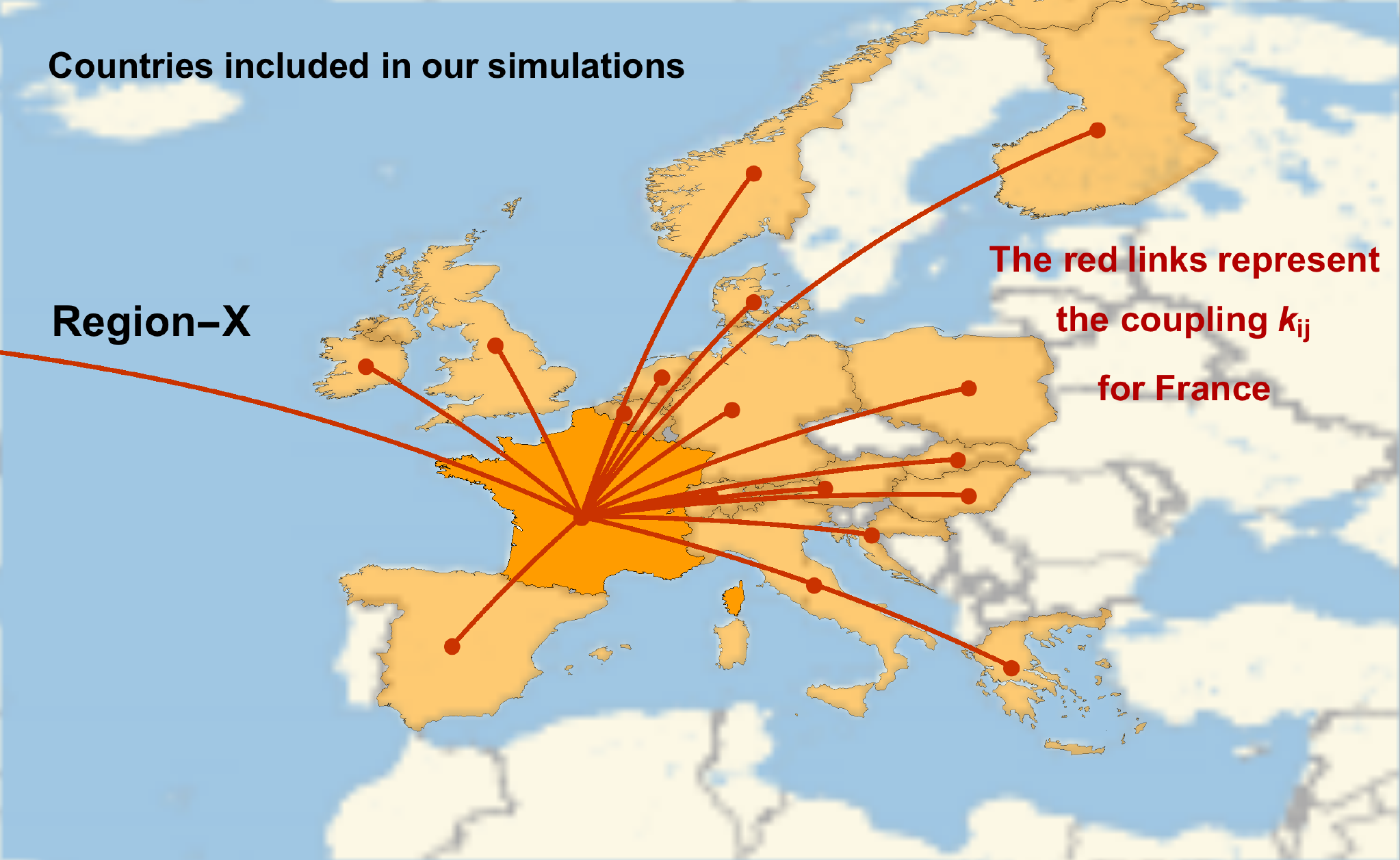}
\end{center}
\caption{Illustration of the connections $k_{ij}$ between, i.e., France and the other countries considered in this study. Each line represents the exchange of infected cases. The line pointing outside the map represents the connection with Region-X, whose role is explained in the main text.} \label{fig:Method}
\end{figure}

The second term of the right-hand side in \eqref{pandemic-eq}, first introduced in \cite{Cacciapaglia:2020mjf}, is a source-term that takes into account human interaction across different regions of the world. Here, $n_{mi}$ is the population of region-$i$ in millions and $k_{ij}$  represents the number of reciprocal travellers per week from region $i$ to region $j$ and vice-versa in units of million people.
For a single country, i.e. France, we illustrate diagrammatically  the connections given by the $k_{ij}$ couplings in Fig.~\ref{fig:Method}.
We also consider an extra-source of infection modelled as a new region that we call Region-X ($i=0$). We can interpret this region in various ways: for instance, this may represent an inflow of infections coming from outside of the regions of the world included in the simulation or, alternatively, Region-X may represent the effect of local hotspots of infections. 
Of course, it could also be a combination of the two effects.
  
\section{Methodology}
  
 To simulate the European second wave, we use as input parameters the values of $\gamma_i$ and $a_i$ stemming from the first wave.   
 Predicting these parameters for the second wave is hard, as shown for instance in Ref.~\cite{2006.05081} via a \emph{stochastic SEIR} model where very large fluctuations are found. This is one of the reasons why we choose for our simulations the parameters coming from the first wave. Additionally this choice has the advantage of endowing us with reasonable benchmark values. 
 These parameters depend on social distancing measures enacted by each country during the first wave. The methodology of the fit for $\gamma_i$ and $a_i$ is described in \cite{DellaMorte:2020wlc,Cacciapaglia:2020mjf}. The values are reported in the first three columns of Table~\ref{T1} at 90\% confidence level. For the simulations we used the central values. 
 
 We now move to the interaction across the different European countries encoded in the matrix $k_{ij}$. We generate the entries of the matrix randomly with each value in the interval $10^{-3} - 10^{-2}$ and a flat probability. This translates in a range of 1k to 10k travellers per week across countries. In our earlier work \cite{Cacciapaglia:2020mjf} this interval was shown to be able to  account for the peak delay in between countries. 
 
 As mentioned earlier, we also consider the extra-source of infection  Region-X ($i=0$) with a fixed number of infected cases. This region couples to the different European countries with  randomly generated $k_{0i} = k_{i0}$ in the same range as above. 
 %In the simulations we will also consider cases in which some of the $k_{0i}$ are set to zero or suppressed. 
 To Region-X we can assign different interpretations. One could be that of an extra-European source (say the rest of the world) that still couples to some or all European countries we consider. Another interpretation is that the coupling $k_{0i}$ to Region-X represents an internal source of infection inside the $i$-th region. To provide a sensible value for the initial source, we considered the current number of total infected (5.2 millions) normalised to the world population in millions.  
 
 Specifically, we randomly generate 100 copies of the matrix $k_{ij}$ to be used to repeat the simulation. The initial time of the second wave simulations is the calendar week 25, where we  set the initial values for $\alpha_i = 0$ (while $\alpha_0 = {\rm constant}$). 
We repeat the 100 simulations with the same set of $k_{ij}$ for five cases, where we modify the coupling to Region-X as follows:
 \begin{itemize}
     \item[a)]{We use the randomly generated $k_{0i} = k_{i0}$, in the range $10^{-2} - 10^{-3}$;}
     \item[b)]{We divide the $k_{0i}$ by a factor of ten, implying a 90\% reduction of the interaction with Region-X;}
     \item[c)]{We divide the $k_{0i}$ by a factor of hundred, i.e. a 99\% reduction;}
     \item[d)]{All the $k_{0i}$ are set to zero except one, which we chose to be that of Spain;}
     \item[e)]{All the $k_{0i}$ are set to zero except the ones for Croatia, Greece, Slovakia, Spain and Switzerland.}
 \end{itemize}
 We consider the latter case e) as the most realistic, as the five chosen countries already show signs of a second wave as of calendar week 30. 
For each of these five cases, we average over the 100 simulated matrices $k_{ij}$ to extract the location of the peak of the newly infected cases for the second wave per each country. The results are summarised in last five columns of Table~\ref{T1} with the errors representing one standard deviation. The time is given in 2020 calendar weeks. 
  
\begin{center}
\begin{table*}[th!]
\begin{tabular}{ ||p{1.8cm}|p{2.cm}|p{2.cm}||p{1.8cm}|p{1.8cm}|p{1.8cm}|p{1.8cm}|p{1.8cm}||}
\hline
\multicolumn{3}{||c||}{First wave parameters} & \multicolumn{5}{c||}{Second wave simulations: peak timing (calendar weeks 2020)}\\
\hline
 & \multicolumn{1}{c|}{$a$} & \multicolumn{1}{c||}{$\gamma$} & \multicolumn{1}{c|}{case a} & \multicolumn{1}{c|}{case b} & \multicolumn{1}{c|}{case c} & \multicolumn{1}{c|}{case d} &
 \multicolumn{1}{c||}{case e} \\
\hline
Austria & $7.463\pm0.007$ & $0.99\pm0.025$ & $30.4\pm0.5$ & $32.4\pm0.5$ & $34.7\pm0.6$ & $38.4\pm0.9$ &
$34.2\pm0.4$ \\
Belgium & $8.53\pm0.02$ & $0.55\pm0.02$ & $34.8\pm0.7$ & $38.2\pm0.7$ & $41.6\pm0.6$ & $43.9\pm1.2$ &
$38.6\pm0.5$ \\ 
Croatia & $6.268\pm0.007$ & $0.71\pm0.02$ & $30.9\pm0.6$ & $33.6\pm0.7$ & $36.6\pm0.7$ & $39.9\pm1.1$ &
$30.9\pm0.7$ \\
Denmark & $7.667\pm0.008$ & $0.40\pm0.01$ & $35.6\pm0.6$ & $39.3\pm0.5$ & $42.8\pm0.5$ & $44.7\pm1.2$ &
$39.4\pm0.6$ \\
Finland & $7.190\pm0.005$ & $0.385\pm0.006$ & $35.5\pm0.7$ & $39.2\pm0.5$ & $42.7\pm0.5$ & $44.5\pm1.2$ &
$39.1\pm0.6$ \\
France & $7.711\pm0.006$ & $0.58\pm0.012$ & $36.2\pm0.6$ & $39.5\pm0.6$ & $42.9\pm0.5$ & $45.2\pm1.2$ &
$39.9\pm0.5$\\
Germany & $7.679\pm0.007$ & $0.62\pm0.02$ & $35.9\pm0.6$ & $39.2\pm0.5$ & $42.5\pm0.4$ & $45.1\pm1.2$ &
$39.8\pm0.5$ \\
Greece & $5.537\pm0.009$ & $0.57\pm0.02$ & $32.5\pm0.6$ & $35.8\pm0.5$ & $39.2\pm0.5$ & $41.8\pm1.2$ &
$32.6\pm0.7$\\
Hungary & $6.022\pm0.009$ & $0.47\pm0.01$ & $34.0\pm0.6$ & $37.5\pm0.5$ & $41.0\pm0.5$ & $43.1\pm1.1$ &
$37.6\pm0.6$\\
Ireland & $8.580\pm0.008$ & $0.60\pm0.02$ & $33.0\pm0.6$ & $36.0\pm0.7$ & $39.4\pm0.6$ & $42.4\pm1.2$ &
$37.0\pm0.5$\\
Italy & $8.304\pm0.004$ & $0.429\pm0.008$ & $39.3\pm0.7$ & $43.0\pm0.5$ & $46.4\pm0.5$ & $48.2\pm1.1$ &
$42.8\pm0.5$\\
Netherlands & $7.904\pm0.005$ & $0.525\pm0.008$ & $35.1\pm0.7$ & $38.6\pm0.6$ & $42.1\pm0.5$ & $44.5\pm1.2$ &
$39.0\pm0.5$\\
Norway & $7.356\pm0.006$ & $0.58\pm0.02$ & $32.7\pm0.6$ & $35.8\pm0.7$ & $39.2\pm0.6$ & $42.0\pm1.1$ &
$36.7\pm0.5$ \\
Poland & $7.13\pm0.03$ & $0.182\pm0.007$ & $46.3\pm0.6$ & $49.9\pm0.6$ & $53.2\pm0.6$ & $54.5\pm1.3$ &
$49.4\pm0.8$ \\
Slovakia & $5.67\pm0.02$ & $0.59\pm0.04$ & $31.7\pm0.7$ & $34.8\pm0.7$ & $38.2\pm0.6$ & $40.9\pm1.1$ &
$31.7\pm0.7$ \\
Spain & $8.747\pm0.008$ & $0.46\pm0.01$ & $38.2\pm0.7$ & $41.9\pm0.5$ & $45.3\pm0.5$ & $38.7\pm1.1$ &
$38.5\pm0.9$ \\
Switzerland & $8.196\pm0.003$ & $0.72\pm0.01$ & $32.3\pm0.6$ & $35.0\pm0.7$ & $38.1\pm0.7$ & $41.5\pm1.1$ &
$32.3\pm0.6$ \\
UK & $8.353\pm0.007$ & $0.368\pm0.007$ & $41.2\pm0.7$ & $44.9\pm0.5$ & $48.2\pm0.6$ & $49.8\pm1.2$ &
$44.6\pm0.6$ \\
 \hline
 \end{tabular}
 \caption{Left block: parameters fitted from the first wave. Right block: median peak time of the second wave for the 5 typologies (cases a--e) we use in the simulations, with 1 standard deviation. The median and error only take into account the 100 simulations, differing by randomly generated matrices   $k_{ij}$.}
 \label{T1}
 \end{table*}
 \end{center}

   \section{Results}

\begin{figure}[tbh!]
\begin{center}
\includegraphics[width=8cm]{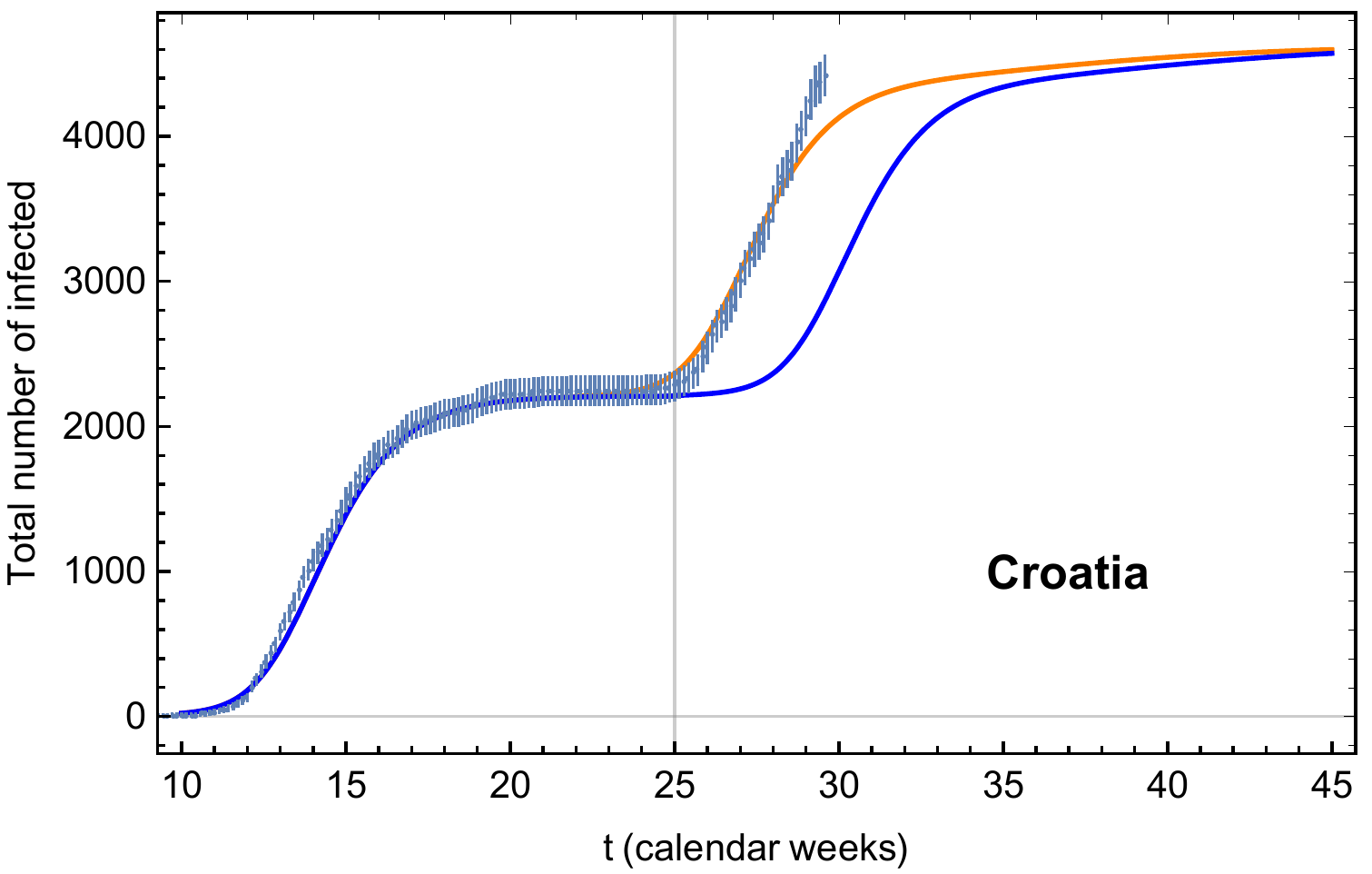} 
\end{center} 
\caption{Croatian number of total infected cases (not normalised per million) with respect to two theoretical curves. The blue one is the result of the simulation as described in the text. The orange curve is constructed by artificially shifting the second wave by three weeks, in order to match the timing in the data.} \label{fig:Croatia}
\end{figure}

We first discuss the results for the simulations in case e), which are more realistic \emph{vis {\`a} vis} the current situation in Europe, as of week 30.
As a test, in Fig.~\ref{fig:Croatia} we show the outcome for Croatia, where we also include the first wave from the fit, compared to the actual data points (from {\tt worldometers.info}). 
The blue curve is the result of one of the 100 case e) simulations, while the orange curve contains the same simulation shifted back by three weeks. The shift could be achieved by increasing the coupling $k_{i0}$ for Croatia by about one order of magnitude (i.e. of the order of $0.1$), to reflect the presence of hotspots inside the country. This is already observable from the data starting at week 25. The figure clearly shows that the infection rate $\gamma$ for the second wave is very close to that of the first wave and that the simulation provides a reasonable understanding of the second wave dynamics. For Croatia we also observe, however, that the total number of infected cases for the second wave is higher than for the first wave. It would be interesting to learn, from future data, whether this worrisome trend is followed by other European countries. The figure demonstrates that the result of our simple simulation can be tuned to reproduce the beginning of the second wave already observed in some countries. This fine tuning is, however, beyond the scope of this work.  
 
 \begin{figure*}[tbh!]
\begin{center}
\includegraphics[width=17cm]{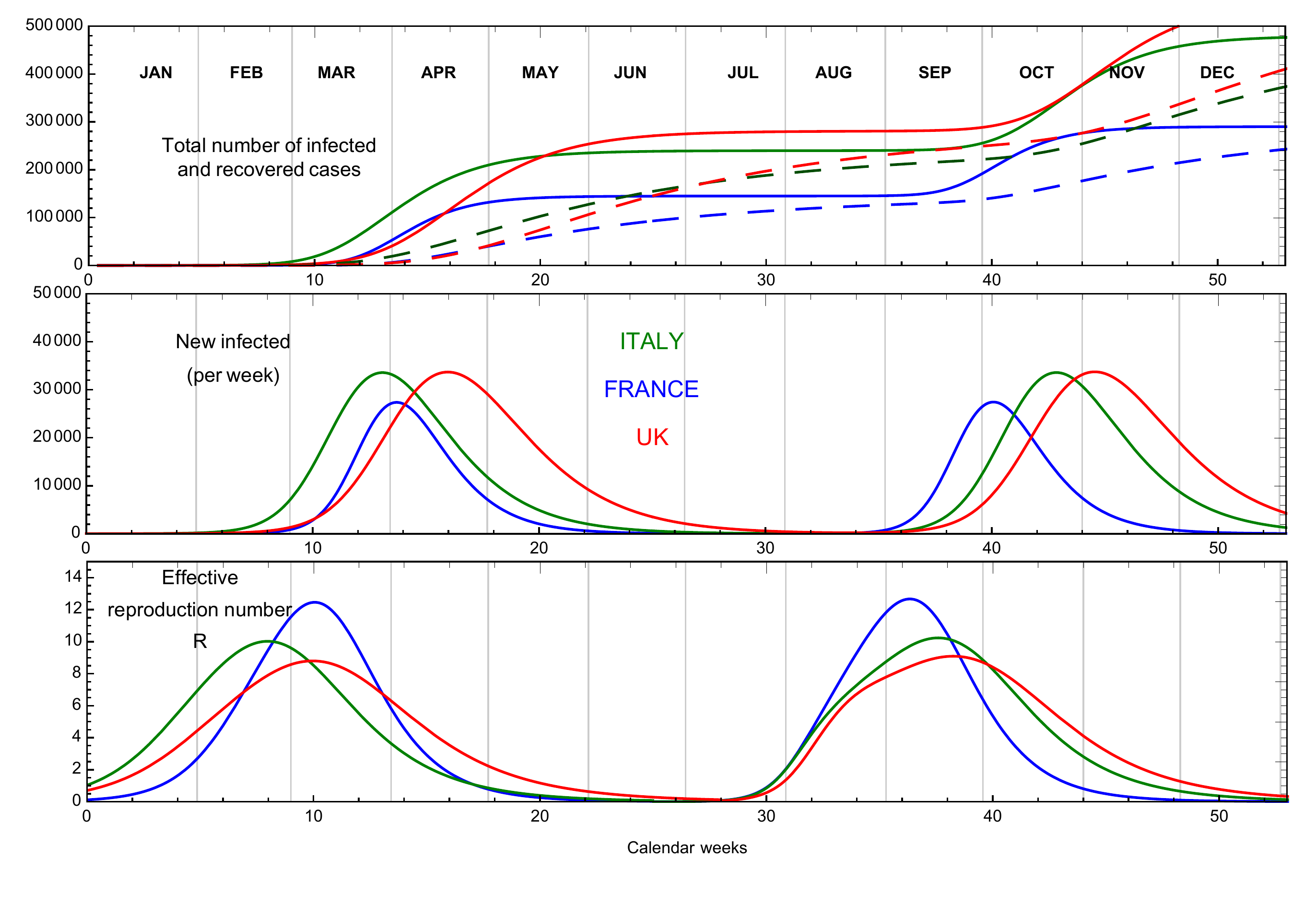}
\end{center}
\caption{Result of case e) for France, Italy and the UK. We show the time evolution of the total number of infected (solid) and recovered (dashed) cases in the top panel, the new infected in the central panel and the derived reproduction rate $R_0$ in the bottom panel. The number of cases refer to the total population of the countries. The shown solutions have a peak position close to the average value from the 100 simulations.} \label{fig:FraItaUK}
\end{figure*} 

\begin{figure*}[tbh!]
\begin{center}
\includegraphics[width=17cm]{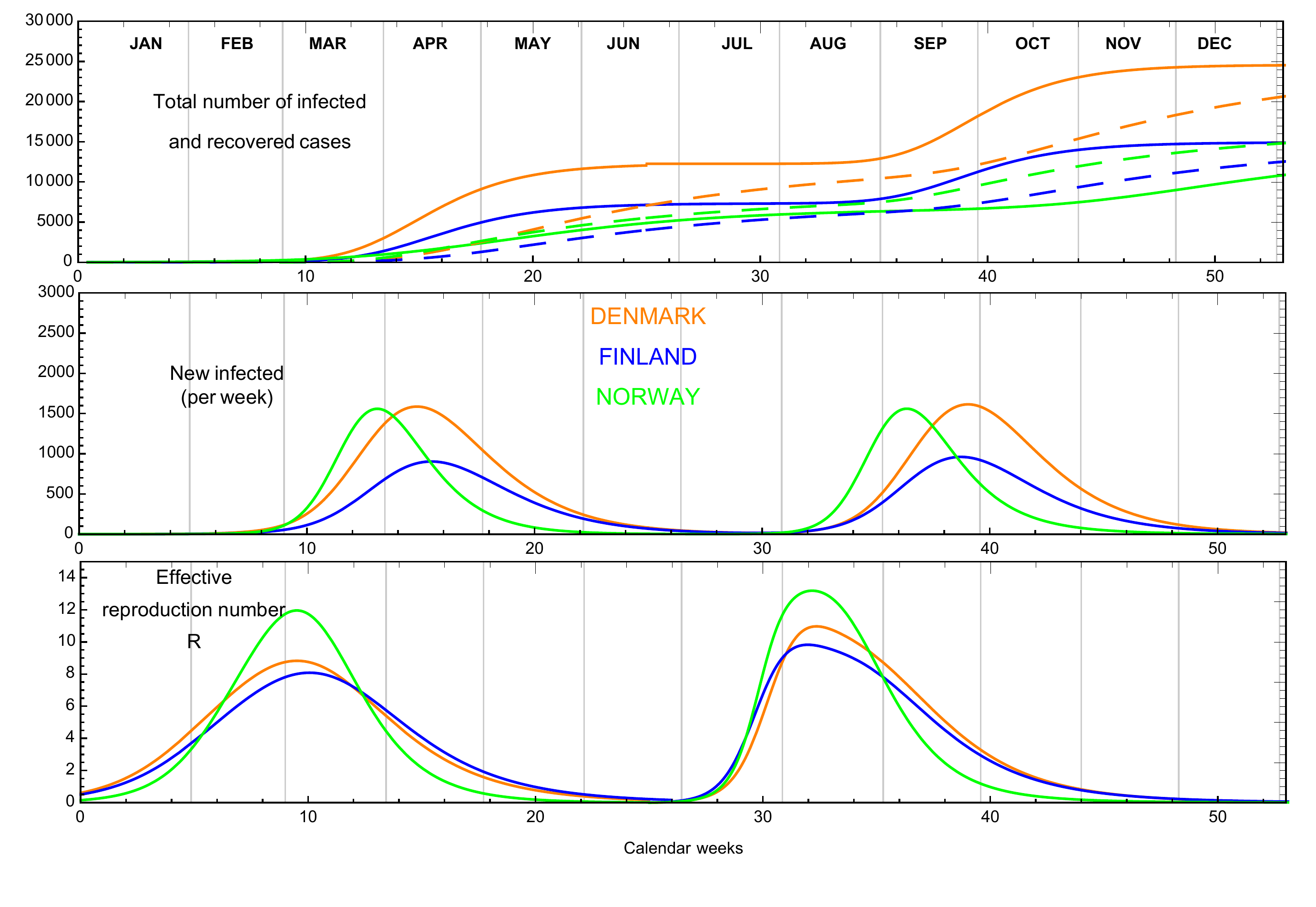}
\end{center}
\caption{Same as Fig.~\ref{fig:FraItaUK} for Denmark, Finland and Norway.} \label{fig:Nordic}
\end{figure*} 
 
As an example of our results for other countries, we show in Fig.~\ref{fig:FraItaUK} the epidemic dynamics of the first and second wave for three representatives:  Italy, France and the UK. The top panel shows the number of infected cases (solid lines) not normalised per million as well as the number of recovered cases (dashed curves). The central panel shows the number of new infected cases while the lower panel displays an estimate for the effective reproduction rate $R$. We also show the results for some of the Nordic countries, i.e. Denmark, Norway and Finland, in Fig.~\ref{fig:Nordic}. 
The number of recovered cases $\mathcal{R}(t)$ is calculated by solving the following SIR-inspired equation~\cite{Cacciapaglia:2020mjf}:
 \begin{equation}
\frac{d \mathcal{R}}{d t} = \epsilon \left( e^{\alpha (t)} - \mathcal{R}(t) \right)\,,
 \end{equation}
 where we fix the recovery rate $\epsilon=0.1$ in the numerical solutions.
 The effective reproduction rate $R$ is estimated by computing the ratio of the new infected cases over the new recoveries within the susceptible population, from the theoretical model. The susceptible population is here defined as the total number of people infected at late time for the first and second waves independently. A more accurate result could be obtained using the generalised eRG approach of Ref.~\cite{morte2020renormalisation}, at the expense of introducing more parameters. 
 The plots are obtained using the simulations for case e). The height of the second wave peaks are the same as for the first wave because we used the same $\gamma$'s and $a$'s stemming from the first wave fit. One could allow for variations of these values, however the qualitative temporal picture of our results would remain similar.

\begin{figure}[tb]
\begin{center}
\includegraphics[width=8cm]{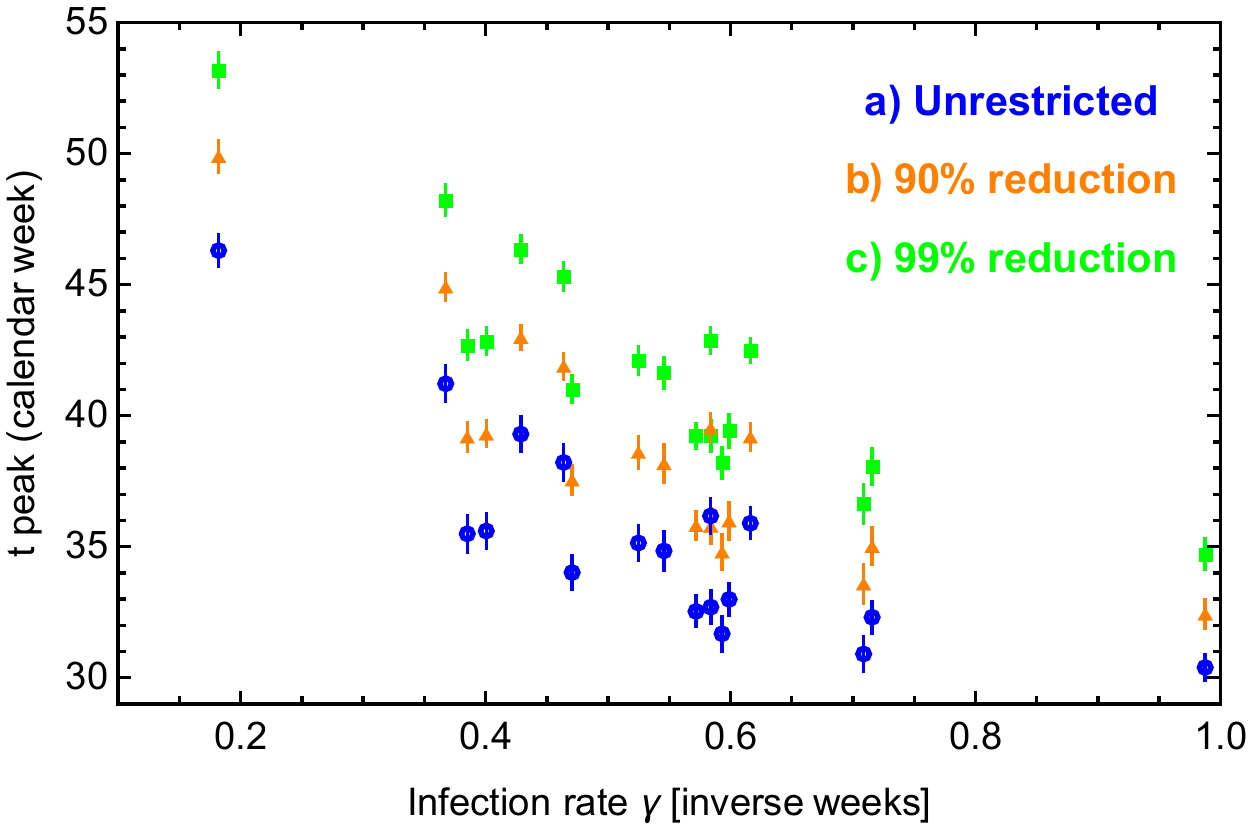}
\end{center}
\caption{Peak time, in calendar weeks, versus the infection rate $\gamma$ for cases a), b) and c).} \label{fig:tpeakvsgamma}
\end{figure}

To study the dependence of the peak timing on $k_{ij}$, $\gamma_i$ and $a_i$, we can use the results from cases a), b) and c) from Table~\ref{T1}, as visualised in Fig.~\ref{fig:tpeakvsgamma}. Here we show the average peak time in calendar weeks versus $\gamma$ for all the countries in this study.
Comparing the results in each set of simulations, we discover a clear correlation between the timing of the peak and the infection rate $\gamma_i$ of each country.  The higher is the infection rate the sooner the peak is reached. Furthermore, comparing the results for the three cases, we show that reducing the coupling with Region-X systematically delays the peaks, in accordance with results reported in \cite{Cacciapaglia:2020mjf}.  Quantitatively a reduction of a factor ten in the coupling to Region-X delays the peaks by about three weeks. We recall that, following the possible interpretations of Region-X, a reduction of the couplings to this region can be seen as the effect of travel bans and/or better control of local hotspots. Overall the peak timing ranges from end of July 2020 to beginning 2021. 
We did not find any correlation between the peak timing and the value of $a_i$ across the countries we studied.
%Case d shows how the spreading dynamics occurs if only one of the European countries, say Spain, is connected to Region-X and therefore acts as European hotspot.  

\begin{figure}[tb]
\begin{center}
\includegraphics[width=8cm]{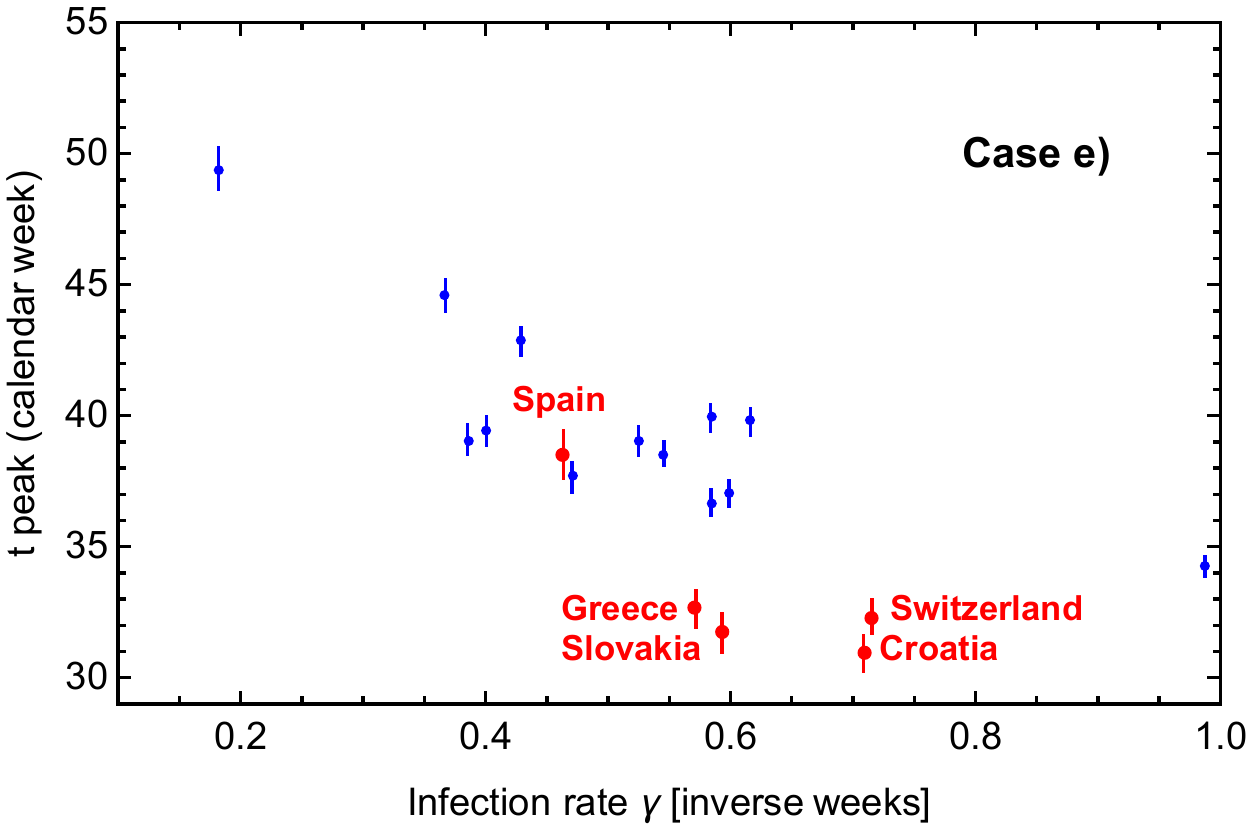} %\hspace{0.5cm}
\end{center}
\caption{Peak time, in calendar weeks, versus the infection rate $\gamma$  for the case e) simulations, with the countries coupled to Region-X highlighted in red. The errors are one standard deviation on the statistics given by the 100 repetitions, as described in the text.}
%\caption{Peak time, in calendar weeks, versus the infection rate $\gamma$ (left) and the log of the final total cases per million $a$ (right) for the case e simulations. The errors are one standard deviation on the statistics given by the 100 repetitions, as described in the text.} 
\label{fig:simul5}
\end{figure}

The results of cases d) and e), where only a few countries act as hotspots, as summarised in Table~\ref{T1}, show a common feature: the peak timing of the hotspot countries is essentially the same we found in the unrestricted case a), as stemming from the $k_{0i}$ values, while the peak timing for the other countries is substantially delayed. The fewer hotspots, 1 as in case d), the more delayed the peak. The results for case e), are shown in Fig.~\ref{fig:simul5}, with the hotspot countries highlighted in red.
It should be clear that the $a$'s and the $\gamma$'s chosen for the simulation can, and will, be different from the first wave values we used. Nevertheless, we expect the dynamics to be still well represented by the framework and that these values give a reasonable indication for the second wave European pandemic.

 %We now move to discuss case e discussed in the section above according to which five European countries act as hotspots. We believe that this is the most realistic case because data show that these five countries (Croatia, Greece, Slovakia, Spain and Switzerland) already display clear signs of starting a second wave at week 30. The results are summarised in  Fig.~\ref{fig:simul5} where in the left panel we show the timing of the peak with respect to the infection rate while in the right panel we show the timing of the peak with respect to the parameter $a$ that represents the log of the final number of infected cases per million of the second wave. It is clear that while the timing is correlated to the infection rate, there is little dependence on $a_i$. 

%%%%%%%%%%%%%%%%%%%%%%%%%%
\section{Discussion and Video Simulation}
\label{sec:5}
We employed the epidemic Renormalisation Group approach to simulate the dynamics of disease transmission and spreading across different European countries for the second COVID-19 wave. Since it  has been demonstrated \cite{morte2020renormalisation} that the framework can be mapped into other compartmental models, our results are sufficiently general. The approach allows to model inter and extra European border control effects while taking into account the impact of social distancing for each country. To reduce the number of unknowns in the simulation, we used the information from the first wave. This information is encoded in the infection rate and the logarithm of the number of total infected cases per each country. Going beyond this hypothesis is straightforward in our approach, but such parameter tuning is not the point of this work. We then performed statistical analyses averaging on different level of  cross Europe interactions and with the rest of the world. The role of the rest of the world and possibly local hotspots  has been attributed to a Region-X, which acts as a source of infection coupled to all or only few European countries.  By calibrating on the current European situation that shows early signs of the second wave, we provided a temporal playbook of the second wave pandemic.  Our results can be employed by governments, financial markets and the industry world to implement local and global measures. 

The main results show that the temporal position of the second wave peak, once started, is rather solid and will occur between July 2020 and January 2021. The precise timing for each country can be controlled via travel and social distancing measures.

In the added material, we also include an animation representing the time evolution of the first and second wave of the European COVID-19 pandemic resulting from one of our simulations close to the average result over 100 simulations for the most realistic case. 
%We also provide a Wolfram Mathematica code, which can be used to tune the simulations to the current data. 
The simplicity of the eRG approach is such that the simulations take only a few seconds on an average personal laptop, thus providing a practical and accurate tool for the understanding of a second (and third, and so on) wave pandemic. The temporal playbook we provide is a useful tool for governments, financial markets, the industries  and individual citizens to prepare in advance and possibly counter the threat of recurring pandemic waves.

\section*{Author contribution}

This work has been designed and performed conjointly and equally by the authors.
  G.C., C.C. and F.S. have equally contributed to the writing of the article.
%F.S. and C.C. analysed the data leading to the fits in Table I, G.C. and C.C. prepared the figures.

\section*{Competing interests}

The authors declare no competing interests.

\end{document}